\documentclass[aps,prl,twocolumn,amsmath,amssymb,
floatfix]{revtex4-1}
\usepackage{tabularx}
\usepackage{bm}
\usepackage{euscript}
\usepackage{epsfig,psfrag,subfigure}
\usepackage{graphicx}
\usepackage{color}
\usepackage{amsfonts}
\usepackage{exscale}
\usepackage{amsbsy}

\usepackage{graphicx}

\def\avg#1{\left\langle#1\right\rangle}

\def\ket#1{\left|#1\right\rangle}

\def\be{\begin{equation}}       \def\ee{\end{equation}}
\def\bea{\begin{eqnarray}}      \def\eea{\end{eqnarray}}
\def\ba{\begin{array} }
\def\ea{\end{array} }
\def\bnum{\begin{enumerate} }
\def\enum{\end{enumerate}}

\def\pa{\partial}
\def\=>{\Rightarrow}
\def\>{\rightarrow}

\def\eye2{\mathbb{I}}

\def\Eq#1{Eq.~(\ref{#1})}
\def\Fig#1{Fig.~\ref{#1}}

\newcommand{\te}{\mathrm{e}}
\newcommand{\eff}{\mathrm{eff}}

\begin{document}
\title{Fermionic magnons, non-Abelian spinons, and spin quantum Hall effect from an exactly solvable spin-1/2 Kitaev model with SU(2) symmetry}
\author{Hong Yao and Dung-Hai Lee}
\affiliation{Department of Physics, University of California, Berkeley, CA 94720, USA}
\affiliation{Materials Sciences Division,
Lawrence Berkeley National Laboratory, Berkeley, CA 94720, USA}
\date{\today}
\begin{abstract}
We introduce an exactly solvable SU(2)-invariant spin-1/2 model with exotic spin excitations. With time reversal symmetry (TRS), the ground state is a spin liquid with gapless or gapped spin-1 but fermionic excitations. When TRS is broken, the resulting spin liquid exhibits deconfined vortex excitations which carry spin-1/2 
and obey non-Abelian statistics. We show that this SU(2) invariant non-Abelian spin liquid exhibits spin quantum Hall effect with quantized spin Hall conductivity $\sigma^s_{xy}=\hbar/2\pi$, and that the spin response is effectively described by the SO(3) level-1 Chern-Simons theory at low energy. We further propose that a SU(2) level-2 Chern-Simons theory is the effective field theory describing the topological structure of the non-Abelian SU(2) invariant spin liquid.
\end{abstract}
\maketitle

Quantum spin liquids are zero temperature exotic states of magnets which exhibit no classical order.
Their existence is often due to strong quantum fluctuations and/or geometric frustrations.
In two dimensions (2D), the notion of quantum spin liquids were first proposed by Anderson in 1973 as the candidate ground state of the spin-1/2 AFM Heisenberg model on the triangular lattice \cite{anderson1973}. In addition, soon after the discovery of the cuprate high temperature superconductors quantum spin liquids are proposed as the root states of the high $T_c$ superconducting states \cite{anderson1987, kivelson1987, lee2006}.
However, up to now there seems no definitive proof of the existence of real spin liquid materials for $D\ge 2$ even though a few promising candidates have emerged recently \cite{lee2008b,balents2010}.  Consequently,
further experimental scrutinies are needed to firmly establish the existence of real spin liquids in $D\ge 2$. Of course, microscopic theoretical models with reliable solution of spin liquid ground states are also desired.

In order for the elementary excitations to carry  well-defined spin quantum number, SU(2) invariance is important. Recently, a number of exactly solvable SU(2) invariant models \cite{moessner2001,lee2003,fujimoto2005,raman2005_sondhi,seidel2009, cano2010} with spin liquid ground states have been proposed. They have the common trait that the low energy/long wave length physics is captured by the short-range resonating valence bond (RVB) \cite{anderson1987,kivelson1987}.

\begin{figure}[b]
\includegraphics[scale=0.24]{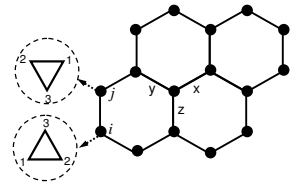}
\caption{The schematic representation of the decorated honeycomb lattice. The triangles are labeled by $i,j$, the sites within each triangle are labeled by $1,2,3$, and the type of inter-triangle bonds is $x$, $y$, or $z$. }
\label{fig:star}
\end{figure}

In this paper, we take a different route to construct a SU(2) invariant model with spin liquid ground states by generalizing the recently discovered Kitaev model \cite{kitaev2006} in a SU(2) symmetric way. Our model [see \Eq{eq:ham}] is a spin-1/2 system on the decorated honeycomb lattice (also known as the star or 3-12 lattice, see \Fig{fig:star}). Our spin model can be reduced to three species of free Majorana fermions coupled to background $Z_2$ gauge field such that the model is exactly solvable, in similar spirit of the original Kitaev model that have only one species of free Majorana fermions.
In the presence of time reversal symmetry (TRS), it exhibits a spin liquid ground state with gapless or gapped spin-1 excitations, depending on parameters in the model. Strikingly, these spin-1 excitations obey fermionic statistics.
When TRS is broken either spontaneously or explicitly, each flavor of Majorana fermions behave like the Bogoliubov quasiparticle of a chiral $p+ip$ superconductor. As a result,
a $Z_2$ vortex binds three Majorana zero modes, protected by the SU(2) symmetry. Due to the existence of
the odd number of Majarana zero modes, the vortex excitations obey non-Abelian statistics, which might find potential application in topological quantum computing \cite{nayak2008}. We further show that the non-Abelian vortex excitations carry spin-1/2 quantum number. As far as we know, this is the first realization of non-Abelian spinon in an exactly solvable SU(2) symmetric spin model. Note that  a SU(2) symmetric wave function with non-Abelian spinon excitations was proposed recently in a spin-1 system\cite{greiter:prl09} but it is not clear which Hamiltonian can realize it as a ground state.

In the SU(2) invariant non-Abelian phase, the system exhibits a spin quantum Hall effect \cite{footnote1}
with quantized spin Hall conductivity $\sigma^s_{xy}=\hbar/2\pi$ (which is twice of that in a $d+id$ superconductor \cite{senthil1999}.) The quantized spin Hall response is shown to be described by
the SO(3) level-1 Chern-Simons gauge theory at low energy.
Furthermore, according to the threefold ground state degeneracy, non-Abelian statistics of vortices, and the SU(2)-invariance of the ground states, we propose that the low energy topological field theory
for the non-Abelian phase
is the SU(2) level-2 Chern-Simons theory \cite{wen1991,fradkin1998}.

{\bf The model:} We consider the following SU(2) invariant Hamiltonian on the lattice shown in \Fig{fig:star}(a):
\bea
H&=&J\sum_i \mathbf{S}^2_i + \sum_{\lambda\textrm{-link}~\avg{ij}} J_\lambda\left[\tau_i^\lambda\tau_j^\lambda\right] \left[\mathbf S_i\cdot \mathbf S_j\right],
\label{eq:ham}
\eea
where $i$, $j$ label the triangles and $\mathbf S_i=\mathbf S_{i,1}+\mathbf S_{i,2}+\mathbf S_{i,3}$ is the total spin of the $i$th triangle ($\mathbf S_{i,\alpha}$ is the spin-1/2 operator on site $\alpha=1,2,3$ of the $i$th triangle.)
The operators $\tau^\lambda_i$ are defined as follows:
$\tau^x_i=2(\mathbf S_{i,1}\cdot \mathbf S_{i,2}+1/4)$, $
\tau^y_i=2(\mathbf S_{i,1}\cdot \mathbf S_{i,3}-\mathbf S_{i,2}\cdot \mathbf S_{i,3})/\sqrt{3}$, and
$\tau^z_i=4\mathbf S_{i,1}\cdot (\mathbf S_{i,2}\times \mathbf S_{i,3})/\sqrt{3}$. The parameter $J$ is the strength of the intra-triangle spin exchange coupling while $J_\lambda~(\lambda=x,y,z)$ describes the inter-triangle couplings on the type-$\lambda$ links.
Because $[\mathbf S^2_i, \mathbf S_j]=0$ and $[\mathbf S^2_i, \tau^{\lambda}_j]=0$, the operator $\mathbf S_i^2$ commutes with the Hamiltonian for all $i$ \cite{wang2010}. As a result, the total spin of each triangle is a good quantum number,  so we can use them to subdivide the Hilbert space.

For each triangle, three spin-1/2's can be decomposed in terms of their total spins: ${\frac12}\otimes{\frac12} \otimes{\frac12}={\frac32}\oplus{\frac12}\oplus{
\frac12}$.
Within the space spanned by ${\frac12}\oplus{\frac12}$ it is straightforward to check that $\vec \tau_i$ satisfy the SU(2) algebra $[\tau^\alpha_i,\tau^\beta_i]= 2i\epsilon^{\alpha\beta\gamma}\tau^\gamma_i$ as well as the Clifford
algebra $\{\tau^\alpha_i,\tau^\beta_i\}=2\delta^{\alpha\beta}$, which implies that $\tau^\lambda_i$ are the Pauli matrices.
In fact the four states in ${\frac12}\oplus{\frac12}$ can be labeled by
$\ket{\sigma^z_i=\pm 1,\tau^z_i=\pm 1}$ with $\vec \sigma_i=2 \vec S_i$ since $[\tau^\alpha_i,S^\beta_i]=0$.
Consequently we may view $\tau^\lambda_i$ as pesudo-spin 1/2 which distinguishes the ``orbital" degree of freedom
within the two degenerate spin-1/2 multiplets.
In the remaining spin-3/2 space, $\tau^z_i=\tau^y_i=0$ because the three spins are totally polarized for $S_i=3/2$; $\tau^\lambda_i$ do not satisfy SU(2) algebra.

When $J\gg J_\lambda$, the ground state and the low lying excited states all lie in the sub-Hilbert space
${\cal L}_0$ where $S_i=1/2$ for all triangles. [It can be shown that the ground state must lie in ${\cal L}_0$ as long as $J>\frac{3}{2}(J_x+J_y+J_z)$.]
For simplicity, we will assume $J\gg J_\lambda$ hereafter which allows us to focus on ${\cal L}_0$. In this sub-Hilbert space, apart from a constant, the Hamiltonian in \Eq{eq:ham}
reduces to
\bea
H=\frac14\sum_{\lambda\textrm{-link}~\avg{ij}}J_\lambda \left[\tau^\lambda_i \tau^\lambda_j\right]\left[\vec \sigma_i\cdot \vec \sigma_j\right],
\label{hl0}
\eea
which seems complicated but is actually exactly solvable as shown below.
To solve the model, we introduce Majorana fermions representations \cite{kitaev2006} for the Pauli matrices  $\sigma^\alpha_i$ and $\tau^\beta_i$ as follows:
\bea
\sigma^\alpha_i \tau^\beta_i =ic^\alpha_i d^\beta_i,~
\sigma^\alpha_i =-\frac{\epsilon^{\alpha\beta\gamma}}{2}ic^\beta_i c^\gamma_i,~
\tau^\alpha_i=-\frac{\epsilon^{\alpha\beta\gamma}}{2} id^\beta_i d^\gamma_i.
\eea
where $\alpha,\beta=x,y,z$ and $c^\alpha_i,~d^\alpha_i$ are Majorana fermions.
As usual, the Majorana fermion representation is over-complete and the following constraint is needed for a physical wave function:
$\ket{\Psi}_\text{phys}$ is in physical Hilbert space iff
\bea
D_i \ket{\Psi}_\text{phys}=\ket{\Psi}_\text{phys}, \forall ~i,
\label{eq:cons}
\eea
where $D_i=-i c^x_ic^y_ic^z_i d^x_i d^y_i d^z_i$. In other words, any state acted by the projection operator $P=\prod_i \left[\frac{1+D_i}{2}\right]$ is a physical state.

In terms of Majorana fermion operators, it is straightforward to rewrite the spin Hamiltonian \Eq{hl0} as follows
\bea
{\cal H}=\sum_{\avg{ij}}J_{ij}  u_{ij} \big[ic^x_ic^x_j+ic^y_ic^y_j+ic^z_ic^z_j\big],
\label{eq:fer_Ham}
\eea
where $ u_{ij}=-id^\lambda_i d^\lambda_j$ and $J_{ij}=J_\lambda/4$ on the type-$\lambda$ ($\lambda=x,y,z$) link $\avg{ij}$. It is clear that $H=P{\cal H} P$.
Because $[ u_{ij}, {\cal H}]=0$, and $[ u_{ij}, u_{i'j'}]=0$,
$ u_{ij}$ are good quantum numbers with eigenvalues $\pm 1$.
It is obvious that \Eq{eq:fer_Ham} is invariant under the following local $Z_2$ gauge transformation
$c^\alpha_i\to \Lambda_ic^\alpha_i$ and $u_{ij}\to \Lambda_i u_{ij}\Lambda_j$, $\Lambda_i=\pm 1$. \Eq{eq:fer_Ham} describes three species of free Majorana fermions coupled with background $Z_2$ gauge field $u_{ij}$.
In addition to the $Z_2$ gauge symmetry \Eq{eq:fer_Ham} has a global SO(3) symmetry which rotate among the three species of Majorana
fermions,
which is the consequence of the SU(2) symmetry of the original spin model.

Lieb's theorem \cite{lieb1994} requires that the ground state lies in the zero flux sector, namely $\phi_p=0$ for every hexagon plaquette $p$ where
$\exp(i\phi_p)\equiv\prod_{\avg{jk}\in p}iu_{jk}$. The zero flux sector is realized by choosing $u_{ij}=1$ with $i(j)\in A(B)$ sublattice.
In the zero flux sector, it is straightforward to show that there are fermionic excitations \cite{footnote2}
that are gapped or  gapless with Dirac-like dispersion, as discussed below.
Besides these fermionic excitations, vortex excitation on plaquette $p$ is created when $\phi_p=\pi$.

{\bf Gapless/gapped spin liquid:} The spectrum of each species of Majorana fermions in \Eq{eq:fer_Ham} can be obtained by a Fourier transform $c^{\alpha}_i=\sqrt{\frac2N}\sum_{\mathbf{k}\in
\frac{\textrm{BZ}}{2}} \left[\te^{i\mathbf{k}\cdot \mathbf{r}_i} \psi_{\alpha,\mathbf{k},a}+\te^{-i\mathbf{k}\cdot \mathbf{r}_i} \psi^\dag_{\alpha,\mathbf{k},a}\right]$,
where $a=A/B$ is the sublattice index of site $i$,  $\{\psi^\dag_{\alpha,\mathbf{k},a}, \psi_{\beta,\mathbf{k}',b}\}=\delta_{\alpha\beta} \delta_{\mathbf{k,k'}}\delta_{ab}$, and the summation is over one-half of the Brillouin zone. The three species have identical dispersion $E_\mathbf{k}=\pm |\sum_\alpha J_\alpha \te^{i \mathbf{k}\cdot \mathbf{e}_\alpha}|$ , which is gapless with Dirac-like dispersion for $J_x+J_y>J_z$ and is gapped otherwise. (Here $\mathbf{e}_\alpha$ label nearest-neighbor vectors.) It is straightforward to calculate the spin-spin correlation with $\avg{S^z_i S^z_j}\sim |\mathbf{r}_i-\mathbf{r}_j|^{-4}$ for $|\mathbf{r}_i-\mathbf{r}_j|\gg 1$ in the gapless phase but in exponential decay for the gapped phase. This is in contrast with the original Kitaev model and its variants studied previously where the spins are uncorrelated beyond nearest neighbors \cite{baskaran2007,tikhonov:arxiv10}.

{\bf Fermionic spin-1 excitations}: In the zero flux sector, what is the nature of fermionic excitations? We show below that these fermionic excitations carry spin-1 quantum number.
For instance, by introducing the complex fermion operators $f_{i,z}=(c^x_i-ic^y_i)/2$, the Hamiltonian \Eq{eq:fer_Ham} can be rewritten as
\bea
{\cal H}=\sum_{\avg{ij}}J_{ij}u_{ij}\left[2\left(if^\dag_{i,z}f_{j,z} -if^\dag_{j,z}f_{i,z}\right) +ic^z_i c^z_j\right],
\eea
which indicates that $f_{i,z}$ are free complex fermions and their fermion number is conserved. The dispersion of $f_z$ complex fermions and $c^z$ Majorana fermions is again $E_\mathbf{k}=\pm |\sum_\alpha J_\alpha \te^{i \mathbf{k}\cdot \mathbf{e}_\alpha}| $ but $\mathbf{k}$ is over the full (half) Brillouin zone for $f_z$ ($c^z$).
Since $S^{z}_i=\sigma^z_i/2 =f^\dag_{i,z}f_{i,z}-1/2$, it is clear that the fermions created by $f^\dag_{i,z}$ carry $S^z=1$, which can be loosely called fermionic magnons with $S^z=1$. The $c^z$ quasi-particles carry $S^z=0$.

In the gapped phase, due to total $f_z$ fermion number conservation (a result of total $S^z$ conservation), the Chern number of $f_z$ can be defined and it is zero. Similarly, a spectral Chern number can also be defined for $c^z$ Majorana fermions and it is also zero. Due to the zero Chern number, a vortex excitation has no fermion zero mode for $f_z$ and no Majorana zero mode for $c^z$; its spin quantum number is zero.

\begin{figure}[b]
\includegraphics[scale=0.35]{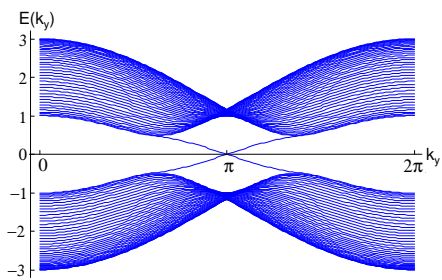}
\caption{The energy spectrum as a function of $k_y$ for $f_z$ fermions on a cylinder periodic in $y$-direction. There is exactly one gapless chiral edge state. The parameters used for both Fig. 2 and Fig. 3 are $J_x=J_y=J_z=1$ and $h=0.3$.}
\label{fig:edge}
\end{figure}

{\bf Non-Abelian spinons}:
A spinon carries spin-1/2 quantum number
and is essentially half of a spin-1 excitation. Deconfined spinon is commonly regarded as a hallmark of a quantum spin liquid. A spinon excitation with $S^z=1/2$ would carry half fermion number of $f_z$. Excitations with half fermion quantum number are known to exist as fermion zero modes in some sort of defect such as domain walls and vortices.
Thus, we study vortex excitations.
Since a fermion zero mode with half fermion number is expected in a vortex with flux $\pi$ when the insulator carries Chern number $\pm 1$.

To obtain a finite Chern number for $f_z$, following Ref. \cite{lee2007}, we consider additional interactions which breaks TRS externally (TRS can be spontaneously broken by ``decorating'' the star lattice \cite{yao2007}):
\bea
H'=h\sum_{\avg{ij}_\alpha \avg{jk}_{\beta}} \epsilon^{\alpha\beta\gamma} \left[\tau^\alpha_i\tau^\gamma_j\tau^\beta_k\right] \left[\mathbf S_i\cdot \mathbf S_k\right],
\label{eq:h}
\eea
where $\avg{ij}_\alpha \avg{jk}_{\beta}$ labels three neighboring sites, which are ordered in a clockwise way within the corresponding hexagon plaquette, forming a triad whose links are $\alpha$ and $\beta$ respectively. In term of fermions,
$H'=\frac h4\sum_{\avg{ij}\avg{jk}} \hat u_{ij}\hat u_{jk} \left(i c^x_i c^x_k +i c^y_i c^y_k +i c^z_i c^z_k\right)=\frac h4\sum_{\avg{ij}\avg{jk}} \hat u_{ij}\hat u_{jk}[ 2(i f^\dag_{i,z} f_{k,z} -i f^\dag_{k,z}f_{i,z}) +i c^z_i c^z_k]$.
This next-nearest-neighbor hopping term generates a mass term that gaps out the Dirac point of $f_z$ fermions
and the resulting insulator of $f_z$ has Chern number $\nu=1$, which can be computed directly and is also indicated by the one gapless chiral edge mode of $f_z$ on a system with boundary, as shown in Fig. \ref{fig:edge}. Since $f_z$ fermions carry $S^z=1$, the finite Chern number $\nu=1$ implies the spin quantum Hall effect $J^{z,i}=\sigma^s_{ij}\pa_j B^z$, where $J^{z,i}$ is the current of spin polarized along the $z$-direction flowing along the $i=x,y$-direction, with quantized spin Hall conductivity $\sigma^s_{xy}=\nu \frac{\hbar^2}{2\pi\hbar}=\hbar/2\pi$.

\begin{figure}[b]
\subfigure[]{
\includegraphics[scale=0.23]{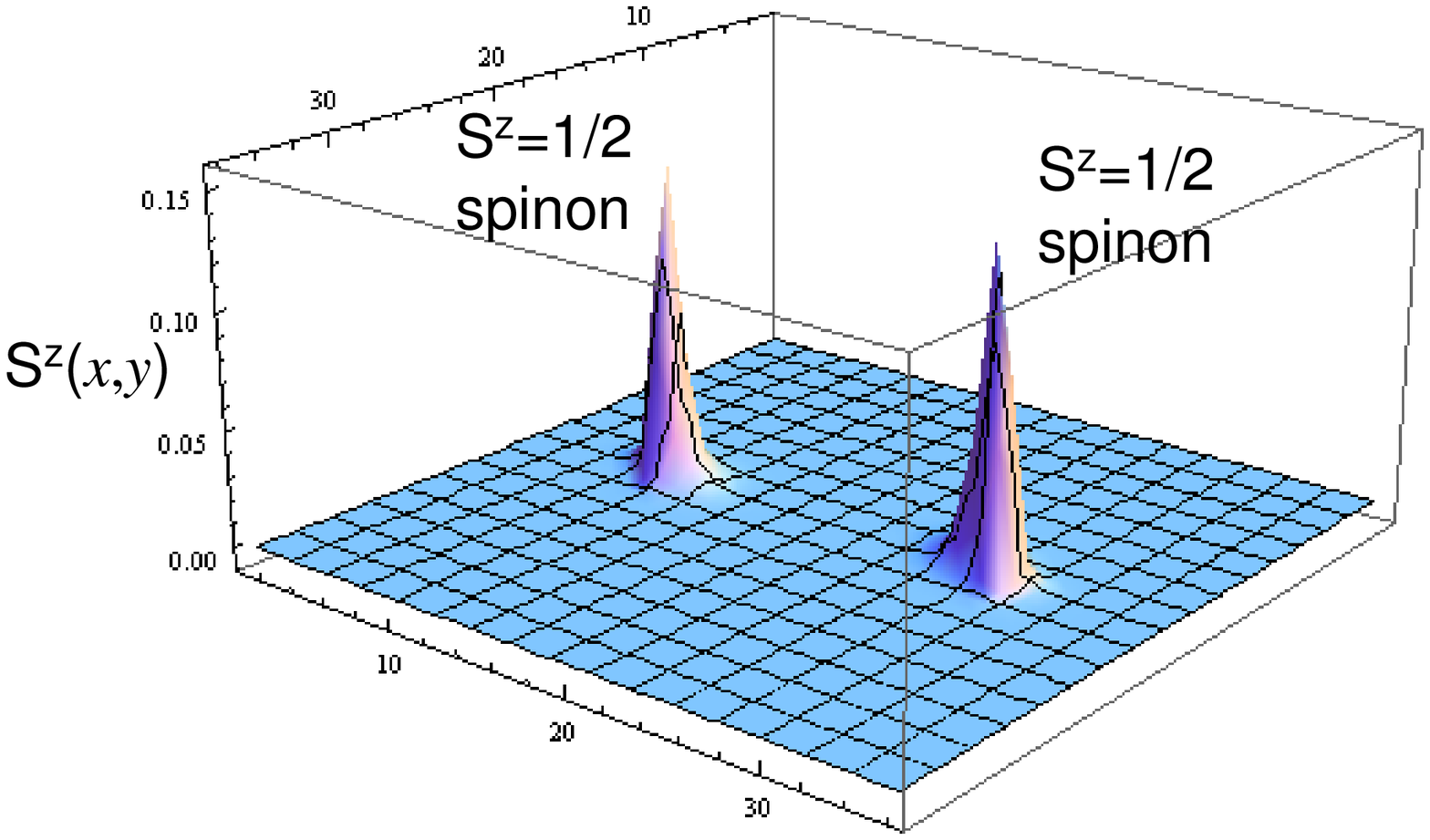}}
\subfigure[]{\includegraphics[scale=0.30]{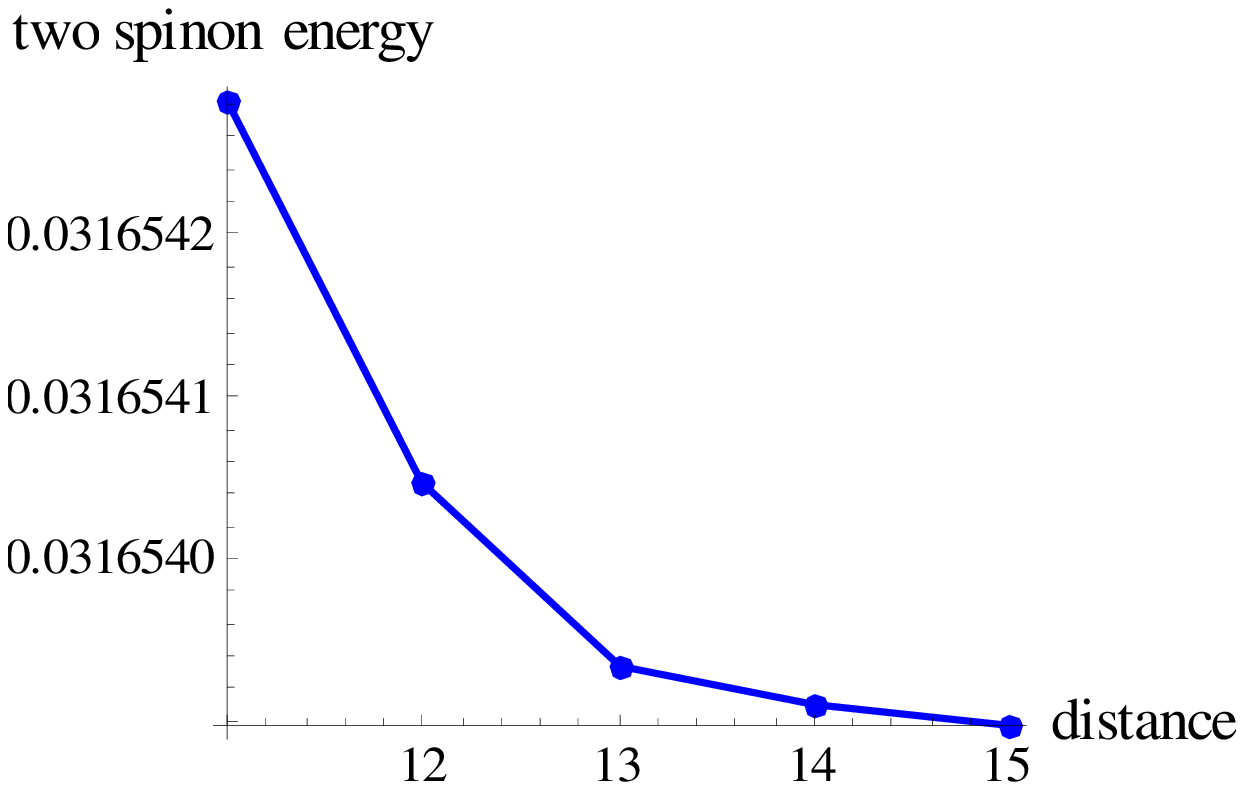}}
\caption{(a) Two localized spinons each with $S^z=1/2$ are realized by creating two vortex excitations. (b) The energy of creating two localized spinons as a function of distance $x$ between then.}
\label{fig:spinon}
\end{figure}

Because of the finite spin Hall conductivity, by inserting a $\pi$ (or equivalently $-\pi$) flux locally on plaquette $p$, namely creating a vortex excitation, the spin is cumulated around the vortex by the amount $S^z=\sigma^s_{xy}\cdot (\pm \pi)=\pm \hbar/2$, implying that the vortex carries spin-1/2 quantum number. The sign ambiguity reflects the facts that there is one zero mode associated with the vortex excitation and that occupying it or not gives rise to $S^z=\pm 1/2$. The spin-1/2 nature of a vortex excitation is further verified numerically. For instance, by creating two well separated vortices in a finite lattice and occupying the zero mode associated with each vortex, we obtain $S^z=1/2$ for each vortex, as shown in Fig. \ref{fig:spinon}(a).

Similarly, the $c^z$ Majorona fermions have spectral Chern number $\pm 1$ and the edge has a gapless chiral Majorana mode. Moreover, there is one $c^z$ Majorana zero mode associated with each vortex excitation \cite{kitaev2006}.
Due to this unpaired Majorana zero mode from $c^z$, the vortex excitations obey non-Abelian statistics. In the following we refer to them as
``non-Abelian spinons'' since they also carry spin-1/2 quantum number. It is now clear that a local vortex excitation actually  binds three Majorana zero modes, which are treated ``on bias'' as a complex fermion zero mode plus a Majorana zero mode above. The three Majorana zero modes will not mix and split because of the SU(2) symmetry.

We further computed the total energy of a pair of vortices as a function of the inter-vortex distance and the result is shown in \Fig{fig:spinon}(b). The fact that the energy decreases with the increasing distance indicates that the vortex-vortex interaction is repulsive and creating two far separated spinons only cost a finite energy. The non-Abelian spinons are deconfined.

{\bf Topological field theory}: A topological phase is generically described by a topological field theory. In the non-Abelian spin liquid discussed above, we expect that the finite spin quantum Hall effect  have a topological field theory description which we will derive below. For simplicity, we assume $J_x=J_y=J_z$ hereafter. Without the $h$ term, the dispersion is gapless with the Dirac point at $\mathbf{K}$. (Note that each species of Majorana fermions has only one Dirac cone due to their Majorana nature.) A finite $h$ term acts like a mass term for the Dirac fermions at $\mathbf{K}$.
In the continuum limit, the low energy physics is then described by the following Euclidean action
\bea
S=\int dt d^2x \bar\psi[i\gamma^\mu \pa_\mu +im ]\psi,
\eea
where $\psi=(\psi_{xA}, \psi_{xB}, \psi_{yA}, \psi_{yB}, \psi_{zA}, \psi_{zB})^T$ and $\bar \psi=i\psi^\dag\gamma^0$. Here $m$ is the mass and
$\gamma^{0,x,y}=\mathbb{I}\otimes\sigma^{z,y,x}$ where $\mathbb{I}$ is a 3 by 3 identity matrix with vector indices $x,y,z$ and $\vec \sigma$ Pauli matrices with sublattice indices.

It is clear that the three species of massive Dirac fermions possess a
global SO(3) symmetry,
which is inherited from the SU(2) symmetry of the original spin model. The continuous SO(3) symmetry allows us to introduce external spin gauge fields $A^a_\mu$, which couple with the spin current $J^{a\mu}=\psi^\dag t^a\gamma^\mu\psi$, where $[t^a]_{bc}=i\epsilon^{abc}$,
in the following way:
\bea
S=\int dtd^2x \bar\psi(t,\mathbf x)\Big[i\gamma^\mu(\pa_\mu+i A^a_\mu t^a)+im\Big]\psi(t,\mathbf x).
\label{eq:gaugecoup}
\eea
By integrating out fermions, we obtain an induced action for $A^a_\mu$, whose lowest-order imaginary part is a topological term:
\bea
S_\eff[A]=\frac{i}{4\pi} \int dtd^2x \epsilon^{\mu\nu\lambda} \Big[A^a_\mu \pa_\nu A^a_\lambda +\frac{1}{3} \epsilon^{abc}A^a_\mu A^b_\nu A^c_\lambda\Big],~~~~~~
\label{tpa}
\eea
which is a SO(3) level-1 Chern-Simons action.
This topological
term describes the spin responses of the system to the external spin gauge fields. Physically, $A^a_0=B^a$ is the external magnetic field.  From \bea
J^{a\mu}=-i\frac{\delta S_\eff[A]}{\delta A^a_\mu}=\frac{1}{2\pi}\epsilon^{\mu\nu\lambda}[\pa_\nu A^a_\lambda+\frac{1}{2}\epsilon^{abc}A^b_\nu A^c_\lambda],~~~~
\eea
we obtain $J^{a i}=\frac1{2\pi}\epsilon^{ij}\pa_j A^a_0=\frac1{2\pi}\epsilon^{ij}\pa_j B^a$ ($i,j=x,y$) which implies a quantized spin Hall response to the gradient of external
magnetic field $A^a_0=B^a$ with quantized spin Hall conductance $\sigma^s_{xy}=\frac{1}{2\pi}$, as expected.

To better understand the topological properties of the non-Abelian spin liquid, an effective topological field theory describing its long-distance and low-energy physics \cite{lee1989,zhang1989,wen1992} would be desired.
To do so, we write $J^{a\mu}=\frac{1}{2\pi}\epsilon^{\mu\nu\lambda}\pa_\nu a^a_\lambda$ so that it satisfies the continuity condition $\pa_\mu J^{a\mu}=0$ automatically. Without derivations, we propose that the following SU(2) level-2 [SU(2)$_2$] Chern-Simons theory \cite{wen1991,fradkin1998}
\bea
S_\eff[a]=i\frac{2}{4\pi}\int dt d^2x \epsilon^{\mu\nu\lambda}\left[a^a_\mu \pa_\nu a^a_\lambda +\frac23 \epsilon^{abc} a^a_\mu a^b_\nu a^c_\lambda \right],~~~~~~
\label{eq:tft}
\eea
is the low-energy effective theory for the non-Abelian spin liquid phase. There are several reasons for such a proposal. First it is natural to use SU(2) as the gauge group; the Majorana fermions has both SO(3) and $Z_2$ symmetries, which combine leading to an effective SU(2) as expected from the spin SU(2) symmetry of the model.
Secondly, as discussed in Ref. \cite{fradkin1998} the edge theory corresponds to the SU(2)$_2$ Chern-Simons theory [\Eq{eq:tft}] is the chiral sector of SU(2)$_2$ Wess-Zumino-Witten model with central charge $c=3/2$, which is consistent with the fact that the edge theory here  has three copies of chiral Majorana modes. Thirdly, the SU(2)$_2$ Chern-Simons theory on a torus is threefold degenerate, which is identical with the degeneracy computed from our lattice model.

{\bf Concluding remarks:} We have shown that the exactly solvable SU(2)-invariant spin-1/2 model on the decorated honeycomb lattice exhibits quantum spin liquid ground states with fermionic magnons or non-Abelian spinons. Interestingly, a recently discovered material, called Iron Acetate \cite{ironacetate}, realizes a spin model on the decorated honeycomb lattice, which adds some hope that our model may be realized in similar family of materials. Moreover, we believe that the model can be potentially realized by loading cold atoms in specially designed optical lattices  under appropriate circumstances \cite{coldatom}.

We sincerely thank Joseph Maciejko, Xiao-Liang Qi, Shinsei Ryu, Ashvin Vishwanath, Zheng-Yu Weng, Shou-Cheng Zhang, and especially Steve Kivelson for helpful discussions. This work is partly supported by DOE grant
DE-AC02-05CH11231.


\begin{thebibliography}{29}
\bibitem{footnote1} Note that spin quantum
  Hall effect is qualitatively different from the quantum spin Hall effect. The
  former is the quantized spin Hall response to the gradient of magnetic fields
  while the latter concerns to electric fields and the spin Hall response is
  not quantized.\BibitemShut {Stop}%
\bibitem{footnote2} As in the original
  Kitaev model, in each flux sector, including the zero flux sector where the
  ground state lies, fermionic excitations with a certain but fixed fermion
  parity are in the physical Hilbert space.
\end{thebibliography}

%

\end{document}